# Crypto-bound identity-verified capability tokens for coordinating distributed AI agents: A proposal

**Srikumar K. Subramanian[1*], Shubhashis Sengupta[1*]**

[1]Accenture Labs

* Correspondence: srikumar.subramanian@accenture.com, shubhashis.sengupta@accenture.com

**Keywords:** Cybersecurity · AI agents · Capability Based Security · Decentralized Identifiers · prompt injection

**Abstract**

The prospect of fully autonomous transactional agents did not appear on the horizon until the advent of high capability language models. With such models, the operational benefits of adaptive task orchestration and independent (but constrained) decision making are tantalizing for enterprises and individuals alike. However, each such agent carries with it a serious attack surface in the form of prompt injection which can compromise any soft "guard rails" that may have been placed in context. The consequences of these attacks include credential ex-filtration which, if left unmitigated, renders the whole category of such agents unusable due to breach of trust. Furthermore, expecting a growth of autonomous agents, a security framework for them would require a form of decentralization to scale. Drawing on the proposed OAuth Agent Authorization Profile and the W3C DID and VC standards, we propose a framework based on the principles of capability based security with decentralized agent identity whereby agents access services based on tokens that are cryptographically bound to the agent's and issuer's identities and specify their scope. Services can validate that the delegation chain only involves scope attenuation before acting on any given token. We show that such a layer that lives outside the language model's context window in a secure module can enable agents to act within enforceable security boundaries.

## INTRODUCTION

The 2025 MIT AI Agent Index documents rapid proliferation across industries, with agent capabilities outpacing the governance frameworks needed to manage them [1]. Garzon et al [2] have written about giving AI agents long lived identities and verifiable credentials based on the W3C protocol proposals including DID. There is wide recognition that agents will require such cryptographic mechanisms in order to ensure their safe, flexible operation across multiple service verticals. [3] Limiting the possible damage when users grant agents access to their persistent services such as Google Drive and Calendar has also been discussed in [4].

In this work, we consider the case where humans delegate tasks they normally perform on their own to a clutch of agents that take actions **on their behalf**. We therefore use the word "agent" here not only to mean a program that can carry out actions in the world, but one that carries out the action **only on behalf of a human**, either directly or transitively. Without such agents, the humans would authenticate themselves with service provider(s) who'd then make available their services within constraints determined by the history of interactions with the person.

The goal(s) for which people seek to interact with a given service provider may not be apparent to the provider – for instance, an aircraft carrier service can get you a seat aboard a flight but not have an idea of what your final destination is and whether you're going there for business or holiday. In the interest of privacy, it would be preferable for such services to not have the full information about the user's intent, lest it bias the transaction or create additional future risks. It is to meet these goals across a number of relevant service providers that agents are seen to be useful.

Access Control Lists (ACLs) often come up in discussions of such security mechanisms in the enterprise context. Therefore we also include an explanation of why capability tokens are better for this scenario compared to ACLs since we anticipate the prolific and potential ephemeral use of agents.

## RELATED WORK

The IETF OAuth Agent Profile Draft proposal describes authorization mechanisms based on capabilities granted to agents, serialized as signed JWT tokens [5]. The JSON structure contains constraints and verifiable grant trail information that can help a service provider receiving the token establish the authenticity and the authority with which the agent is reaching out.

[6] and [7] propose combining Decentralized Identity (DID) and Verifiable Credentials (VC) to enable agent identity. The DID mechanism, with public directory services such as `plc.directory`, can serve to identify agents via their demonstrable control over domains/subdomains. The VC mechanism

with the BBS+ credential subsets enables the presentation of exactly the information required to an agent or service provider, with each verifiable presentation (VP) being de-linked from others to deter profiling. The W3C also has the CG AI agent protocol in the draft stage, to prescribe mechanisms for handshake between AI agents [8].

Several IETF draft proposals have also been brought about, addressing the problem of agent identity and authorization. The Agent Name Service IETF draft proposal also attempts to provide cryptographically verifiable naming identities for agents (AI or otherwise) as an analog to the DNS service [9] (an overview of ANS is available in [10]). The Agent Identity Registry System [11] draft proposes a federated approach to agent identity management anchored to hardware security modules. The Agent Authorization Profile draft [5] proposes OAuth 2.0 extensions to support authorizations granted to agents that are aligned with the capability based security model.

Key PaaS providers AWS, Microsoft and Google have also put forth their own protocol proposals towards agent identity and authorization [12], [13] and [14]. The certificate authority Digicert has outlined a trust architecture combining PKI, code signing and attestation in their industry whitepaper [15].

## PROPOSAL OUTLINE

Recognizing that centralized management of agent identity cannot be expected to be viable at scale, we describe the application of the W3C DID and VC standards along with an OAuth-draft-like capability token model with strictly attenuated and limited delegation as a protocol for enabling agents to act securely on behalf of humans.

In the sections that follow, we first describe why capabilities offer a better security model than ACLs for AI agents. We then use an example scenario in which a user engages a group of specialized agents to help plan a holiday trip and make the necessary reservations. We describe the architecture and the message exchanges that happen as part of this scenario and discuss open design questions pertaining to token revocation.

## IDENTITIES AND ACCESS CONTROL

Determining the scope of access to grant to an agent requires two key techniques – reliably identifying an agent, and for each resource and service, determining what authorities to access the resource that the agent carries.

### Identities

The simplest form of an identity is a public-private key pair associated with the agent, by some trusted party attesting to that association through an appropriate digital signature. While assigning such a key pair to an agent is simple, doing so in such a way that when the agent presents a validation of this association to a service it needs to access, the service can validate the association without a trusted party to call on is non-trivial.

Since capabilities in these scenarios originate with human users, the endpoint services (such as banks for making payments) which the delegate agents need to access, will need to have registered the user's credentials by public key in order to recognize the capability tokens that agents may provide. For payments, an alternative is to have the agent service providers maintain their own registered accounts and charge the user in the end. However, in the latter case, users have less control over (accidental or intended) rogue behaviour of the agents.

[16] presents an approach that encodes information pertaining to an identity in a "DID document" which is a JSON structure. The "ATProto" protocol uses a public directory hosted at https://plc.directory that serves DID documents by id. The trust in the directory therefore partially transfers to the trust in the association between a DID document and an agent that possesses the required private keys to prove control over the ID. The tentative protocol proposed in [8] uses DID documents as a handshake before agents exchange tasks and capabilities.

Internet domain names in combination with secure access provided via HTTPS, using certificates issued by known authorities, can also serve to reliably identify an agent that claims to belong to the domain. If a service runs multiple agents, they can each be placed under a separate subdomains with individual certificates that can help establish that the human or agent communicating with the agent is doing so with the right agent and not an impostor.

The HTTPS protocol's TLS security layer already relies on trusted certificate issuers stored on the browser's end in order to establish credibility of interactions with a particular site. However, "Let's encrypt" [17] has made issuing and managing such certificates cheap/free, which enables secure hosting of services possible in a nearly-decentralized manner, rallying around the domain name mechanism. The central entities here then become the domain name registrars.

### Access control

It is common practice to determine whether an entity is permitted to access a given resource by listing the entities and granted permissions associated with each resource. Such lists are known as "access control lists" or ACLs for short. Given a sufficiently robust identity mechanism, entities can be granted specific access per resource at a granular level.

Figure 1 shows the ACL and the "Capability" view of the same set of permissions governing the interaction between a resource and an agent. Since we're considering open systems with potentially ephemeral agents, the latter view is easier to scale, generalize and to support delegation in a decentralized setup.

| | Kumar | Chiranjib | Tanya | Paramita |
|---|---|---|---|---|
| s3:// bucket | R | R | Denied | RW |
| teams:// channel | Denied | Denied | R | RW |
| OneDrive:// folder | R | R | RW | Denied |

**Figure 1:** ACL (row) versus Capability (column)

## THE COLUMN VIEW: CAPABILITY TOKENS

Flipping from the "row" view to the "column" view, as shown in Figure 1, gives us a way to accommodate the constraints described in the previous section. In this view, each agent presents a token that contains verifiable information about its identity, the authority it carries to access the resource (and possibly other resources), the identity and authorities required of the grantor, along with the various constraints imposed by the grantor on the access including an expiry date for the grant.

The combination of identifying a service to which access has been granted in a unforgeable manner, and the means to access a potentially restricted set of its services is referred to as a "capability". What we call a "capability token" is a signed JSON structure that includes information about the agent's identity, the authority that is signing the capability granted to the agent (which can be the user or another agent based on another chained capability token), the target resource the token is intended for, and the constraining parameters of the granted access to the resource. Agents present such a "bearer token" when accessing a service and the service can independently validate the token and the constraints embedded in it before acting on the request. A resource may further require the bearer to prove that they control the identity mentioned in the token.

1. A user can use their identity to cryptographically sign a time limited grant of permissions to access specific services with specific quotas, for a specific agent identified by its cryptographic credentials retrieved from, say a URL location like `https://agentname.somedomain.com/.well-known/did.json` and have the id verified against a public directory like `https://plc.directory`.
2. Having the token be bound to the identity of the agent means that the agent can be asked to prove that it is indeed the intended grantee. An agent with a security layer that handles its assigned private keys can provide such evidence by a common challenge response mechanism - where the verifier provides a random string and asks the agent to sign it, and verifies the signature using the agent's public key known to it a priori. This implies that the agent must be accessible via specific secure URI end points.
3. All constraints on accessing the resources such as the purpose of the transaction, spending budgets, read/write permissions, delegation permissions and so on can be encoded as data in the signed document.
4. A service to which the agent presents the token can independently validate the agent's identity, the user's identity, and check whether the ask of the agent conforms to the constraints laid out in the included token data.
5. When an agent wishes to delegate part of its task to another agent, it can grant it such a capability token that's tied to its own grant, so the capabilities form a chain. A service can verify the delegation chain by ensuring that grants only attenuate the constraints and not amplify them.[1]
6. If this capability granting layer and the involved credentials are maintained in a secure software module, perhaps backed by hardware based security, then exfiltration of credentials via LLM context window can be made impossible by construction since no private key is accessible in any context window and the module will only provide the encrypted and signed results via its interface.

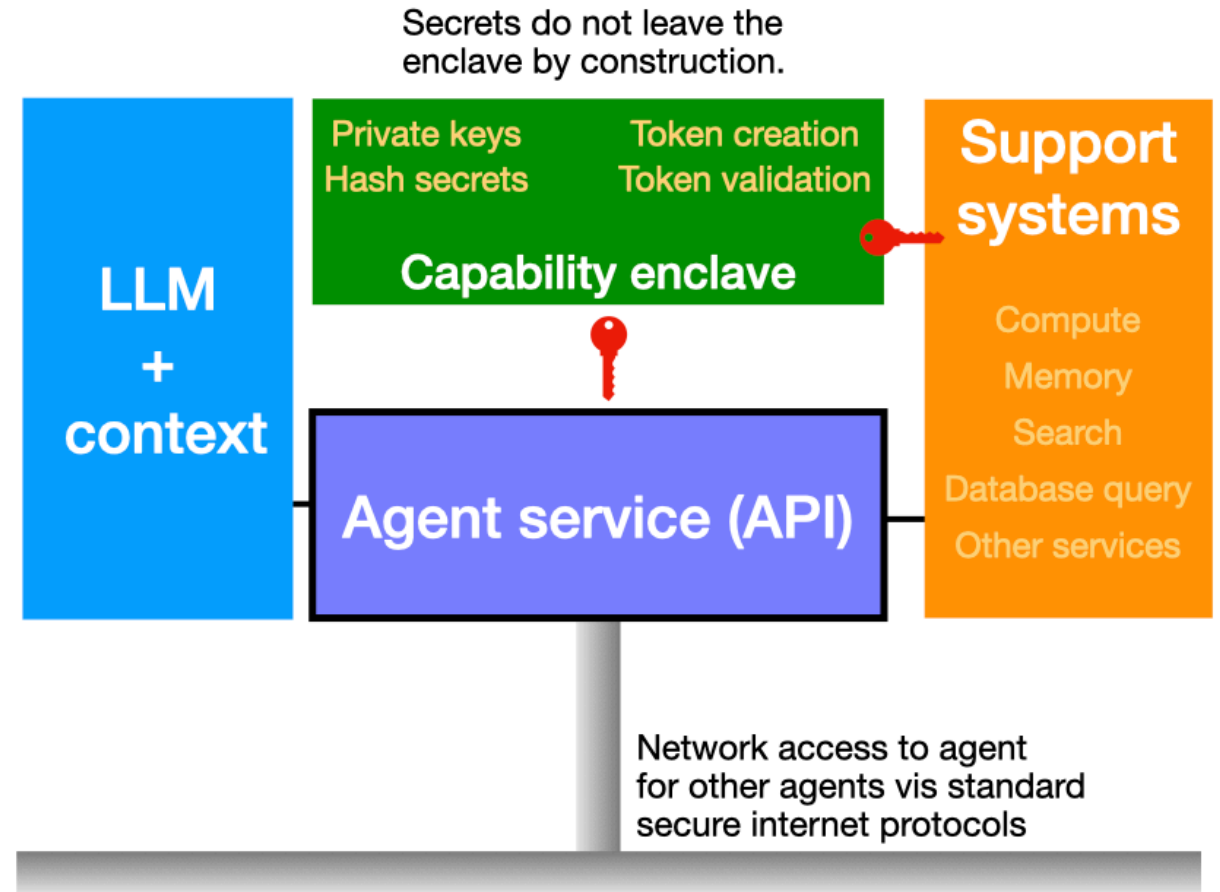

**Figure 2:** Agent service internal structure.

7. AI agents will not only need to access user-aware services on their behalf, but will also need to access tools along the way to accomplish their goals. Capabilities generalize beyond web-hosted services to such tooling where the service layer can validate a capability before granting access to the resource. For example, access to a local file system can be gated by a layer that ensures that an agent can only access a directory or its sub-directories. Access to tools can also be similarly gated, but the tools must themselves use only capability aware system facilities. This is becoming easier to orchestrate in cloud environments via generic capability

[1] Determination of whether the attenuation constraint holds is domain-specific.

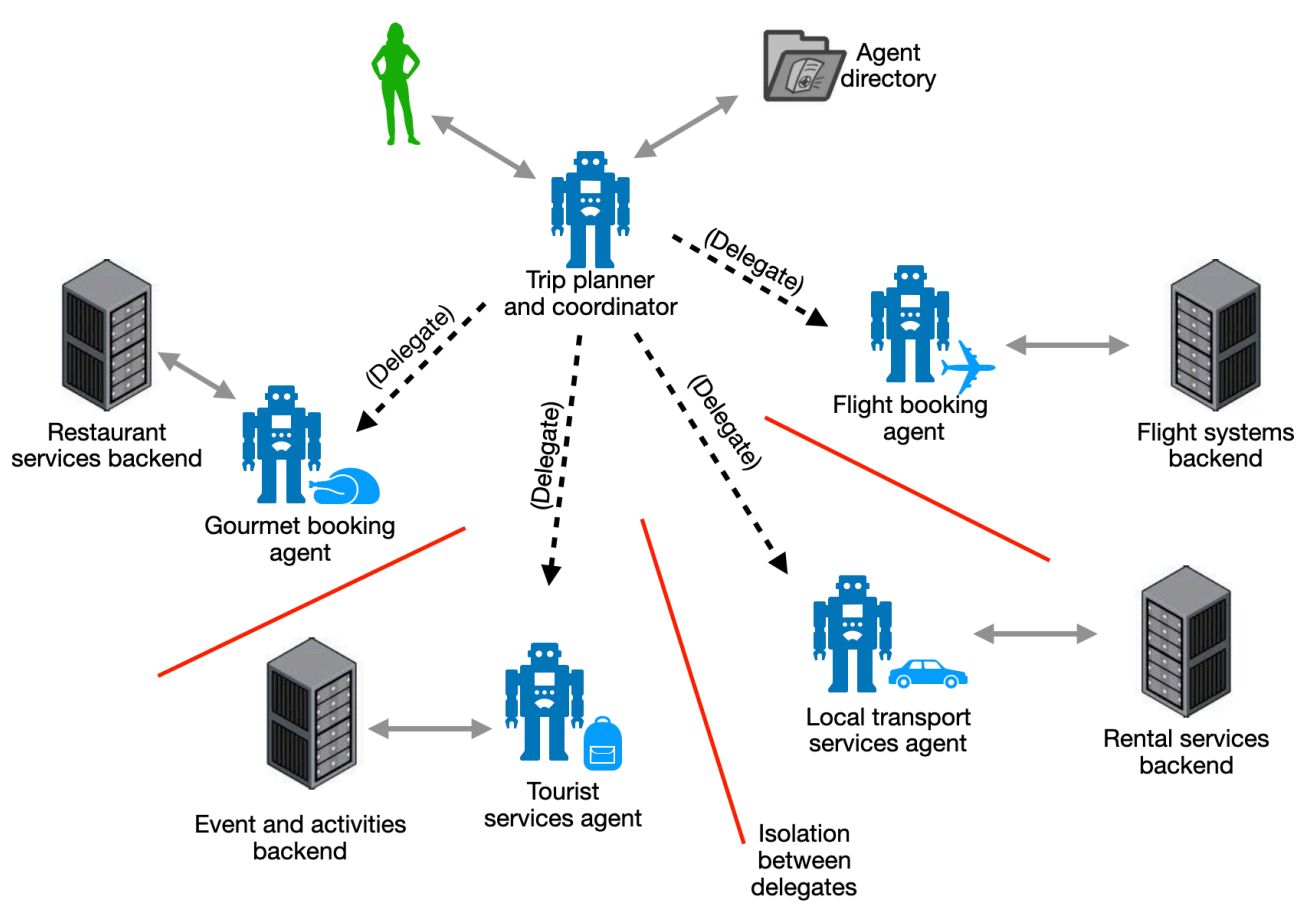

**Figure 3:** Agents collaborating on planning a user's holiday trip, showing the delegation chain.

aware interfaces such as the Web Assembly System Interface [18].

## ILLUSTRATED EXAMPLE

Here, we look at a simplified scenario that serves to illustrate the use of this mechanism for a group of collaborating agents. The agents that participate in the scenario are shown in Figure 3.

Consider a person residing in LA who is looking to plan for a holiday in Paris. The planning activity requires planning for travel by flight, hotel stay for the duration of the holiday in Paris, local transport while in Paris (if the person does not wish to cycle the city), reservations at restaurants of culinary interest to them and bookings to visit various attractions. These are separate concerns and it is to be expected that an agent responsible for booking local transportation does not know anything about other details like hotel stay or tourist activities. Therefore in this scenario, it is preferable to have individual agents plan and perform these activities on behalf of the user, coordinated by a master agent with whom the user shares the trip details.

Here are the steps of the flow -

1. User gives trip parameters ("Los Angeles to Paris trip for 2") and constraints ("budget of USD 5000") to the "Atlas" coordinator agent. This results in a capability token given to Atlas that contains the stated constraints, is signed by the user's private key, and allows the agent to delegate parts of its task to other agents.
2. Atlas decides to get the "Skyway" agent to search for and reserve flights suitable for the trip. It grants Skyway the authority to block flights and gives a reduced budget of USD 1500 for blocking tickets. This also permits Skyway to search for flights and respond to Atlas which presents it to the user for their choice. Once the user picks the flight, Skyway blocks the tickets for the user.
3. Atlas similarly proceeds with the "Haven" agent for booking hotels, "Gourmand" for reserving tables at various restaurants, and the "Roadster" agent for booking local transportation such as rental cars and train tickets, and so on.
4. Once each agent has completed their holds, Atlas presents the entire plan in a consolidated manner to the user and asks for permission to book the trip. Recall that so far, the user has only granted permission to block seats, hotel rooms etc. Not to actually pay for them. At this point, the user may choose to revise the plan, in which case Atlas will revoke the capability tokens granted to the sub-agents and reissue with renewed constraints.
5. Once the user is happy with the plan and grants permission to make the bookings, Atlas will revoke all of the "block" capability tokens granted to the other agents and issue tokens with the permissions to actually spend money on behalf of the user, with more precise amounts given in the token data.
6. The respective agents can now go ahead and make the bookings on behalf of the user and the whole story gets completed.
7. At any point, the user will be able to abort the process by revoking the capability they granted to Atlas as an "emergency stop" measure. When a booking service checks the delegated tokens and follows up the grant chain, they will discover that the grantor's permissions have been since revoked and can deny the service. They can also deny the service if the delegation process failed to attenuate the constraints with each delegation step.

## THREAT MODEL

| Layer | Mechanism | Purpose |
|---|---|---|
| User → Agent | Master capability token | Authorises the coordinator to act on the user's behalf |
| Agent → Agent | Delegated capability tokens (ECDSA-signed JWTs) | Restricts each sub-agent to its declared permissions; prevents privilege escalation |
| Agent → Service | W3C Verifiable Presentation (did:jwk + ECDSA) | Authenticates the agent to external services without exposing raw credentials; wallet policy prevents prompt-injected exfiltration |

**Table 1:** The three layers of security

We assume that humans and agents can prove their control over their respective identities represented as DIDs without a central issuing service. These identities, participating in a signed token chain with no revocations is sufficient authority to permit a service provider to perform the task at hand, and token validity periods are kept short to mitigate timing attacks. We also assume that the core services accessed by the agents on behalf of the user are trusted by the user.

Assets to protect from being exposed to the context windows of LLMs involved in the agents are the various keys required for identity proof, signing, and production of VPs. While it is possible that in some cases the use of high capability LLMs can result in cyber attacks targeted at these assets, the scope of this proposal cannot in general cover the full breadth of such attacks due to variance in implementation that can be expected. It must also not be possible for an LLM to generate valid tokens by guessing secrets used for signing them.

Attackers will be interested in hijacking tokens issued by user to agents for their own use with the target service providers. Such tokens must therefore also be protected from replay attacks. User profiling by attackers can, over time reveal information about the user that can later be used for identity theft. While this cannot be eliminated entirely, any brakes on profiling is desirable, especially in conjunction with short lived tokens. To the extent possible, insiders to service provider infrastructure should not be able to access the guards placed on actions taken given an authority carrying token.

The user uses a service or program to issue tokens and revoke them. This service, especially if accessed over the internet, constitutes an attack surface. It is relatively easy to ensure that the tokens be tamper evident and not observable by passive observers. Exposure to what can get into the context window of LLMs is the prime attack surface being considered in this proposal, since secret exfiltration and constraint modification both compromise the security of the transactions involved.

Threats include attackers claiming a DID that they do not control, modified capability tokens and VPs, prompt injection attacks on any participating agent and any form of privilege alteration by malicious code operating within bounds of valid service or agent providers. From a service provider perspective (such as a payment gateway), it is also important that the party issuing a token with authority to perform an action cannot later deny having issued the token.

## THE CREDENTIAL WALLET

Long lived capability tokens increase the risk of credential exfiltration since cyber attacks triggered by injected prompts cannot generally be ruled out as it depends on actual secure implementation as opposed to theoretical properties of the security protocol. The credential wallet pattern mitigates this risk by removing the ability of an agent to retrieve the raw signed credential and to subsequently be able to generate its own verifiable presentations. The wallet policies can enforce these controls. While it does not eliminate this risk entirely, a careful implementation that accounts for the residual patterns listed in table Table 2 can help substantially.

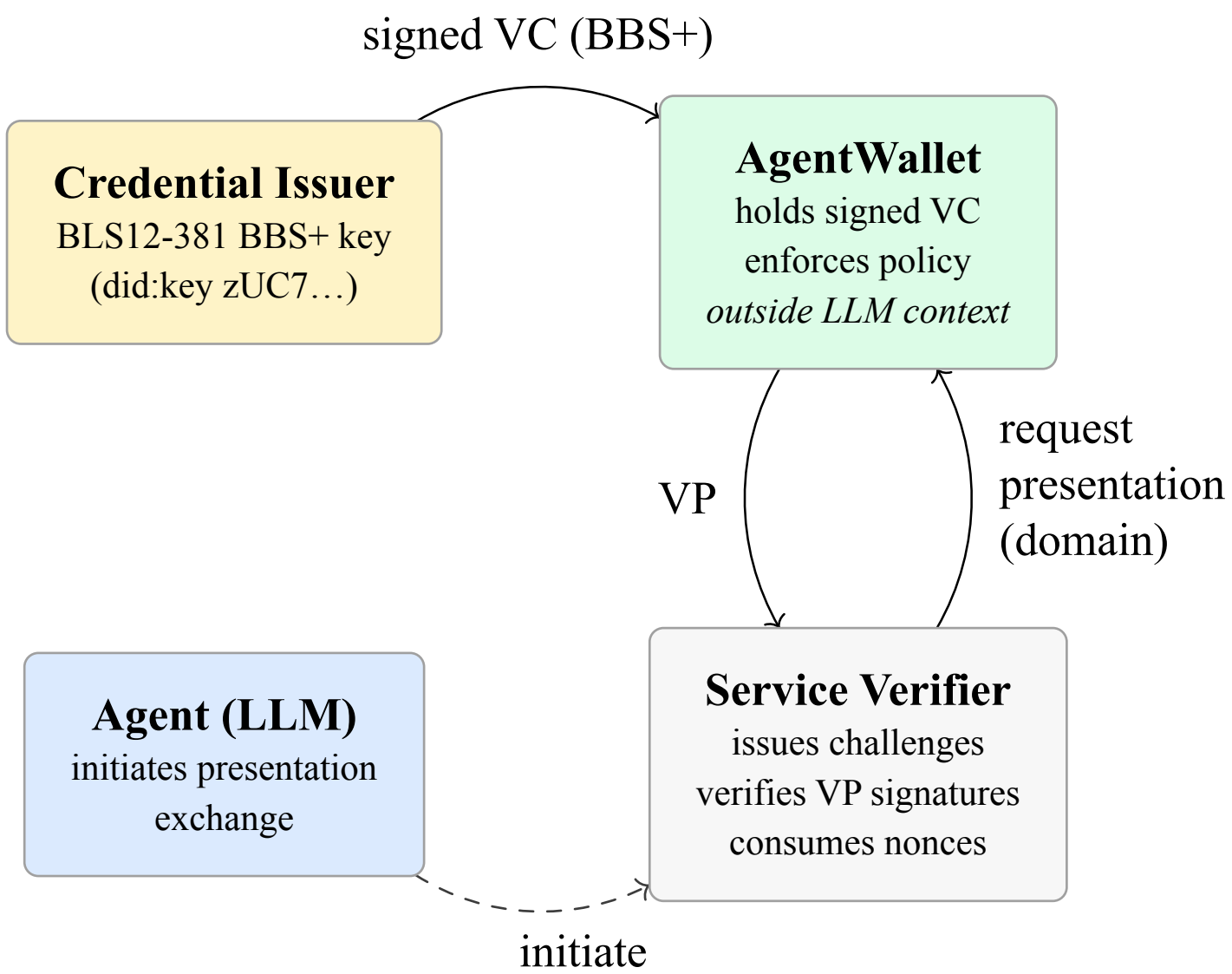

**Figure 4:** Architecture

Table 2 captures risks beyond the following list of recommendations for the wallet implementation.

- Wallet runs as a separate process with typed IPC only.
- Use HMAC with a wallet-held secret instead of a hash like SHA-256.
- Denial messages must be identical for all denials so as not to turn it into an oracle.
- Use BBS-2023 instead of ECDSA based proofs for unlinkability of VPs.

## REVOKING CAPABILITY TOKENS

Revoking a capability token has a fundamental design trade-off to consider. If the token is to be used purely offline – i.e. all aspects of it can be validated by the receiver – then revocation is extremely hard or even impossible. Any intermediary increases transaction latency though.

[19] specifies a "token revocation endpoint" to which a revocation request can be made and is subsequently expected to be propagated to relevant services so that revoked tokens don't end up being used due to lack of information about the revocation. We can make revocation easier by allowing access to a service that can check whether a given token is valid. However this implies that validating a token now incurs a network access step and therefore increases the latency of the transaction completion.

Considering the ubiquity of network access, we think the availability of the validation end point and additional latency is in most cases an acceptable trade-off for the benefit of having the

| Residual risk | Effort | Mitigation |
|---|---|---|
| Aggregate disclosure | Low | Rate-limit requests; require per-presentation user approval |
| VP transits LLM context | Medium | Wallet posts VP directly to verifier; agent receives only pass/fail |
| Challenge acquisition by LLM | Medium | Out-of-band challenge delivery (QR code, second device) |
| UI manipulation of policy | High | Policy changes require out-of-band second-factor confirmation |
| No credential status checking | Medium | Wallet checks W3C Bitstring Status List before signing a VP |
| Unique claim combination re-identifies by content | N/A | Not a cryptographic problem; addressed by minimum-necessary disclosure policy and claim-set diversity analysis |

**Table 2:** Residual risks of the wallet pattern

tokens be revocable at short notice. In the interest of avoiding centralized services as required components, below are some possible design alternatives to consider –

1. Make all tokens short-lived (say 5mins) and force agents to refresh their tokens when they time out. If the refresh fails due to a revocation, the token is observed to be invalid. If transactions can be completed quickly, this can be viable, but it too introduces a latency cost to enable revocation.
2. Make every token include a reference to the parent token and have the service provider walk the parent chain to ensure validity of the entire grant chain. This enables revocation at any sub-tree level, but the latency is now dependent on the depth of the delegation tree.
3. Use a cryptographic accumulator published at a known end point that validators can check against, and include witnesses in the capability along with the accumulator epoch. If a token in the grant chain gets revoked by being removed from the accumulator, the validator will observe that as a validation failure of the supplied capability. This is complex cryptography to manage, but has some useful properties as well.

| Participant | Role |
|---|---|
| User | Human at the browser; trust anchor for the whole session |
| Wallet Module | BBS+ key generation, credential issuance, and `AgentWallet` / `ServiceVerifier` instances — runs in the browser, outside the LLM context |
| Atlas | Coordinator agent; orchestrates booking and issues delegated capabilities |
| Skyway | Flight agent; searches and holds flights under its delegated capability |
| Haven | Hotel agent; searches and reserves hotels under its delegated capability |
| ServiceVerifier | Per-domain verifier object; issues single-use challenges and verifies VPs |

**Table 3:** Participants

## FLOW OF EVENTS

In this section, we discuss the detailed flow of events starting from the initiation of the request with the coordinator agent and the completion of payment towards the bookings made. The main actors are shown in Table 3.

During the initial communication of capabilities between the user and the designated agents, as shown in Figure 5, the wallet module bootstraps in the background preparing the verifiers and getting the required credentials. The wallet is directly controlled by the user. The agents do not have the ability to exchange data with the wallet apart from instructing it to communicate with the respective parties.

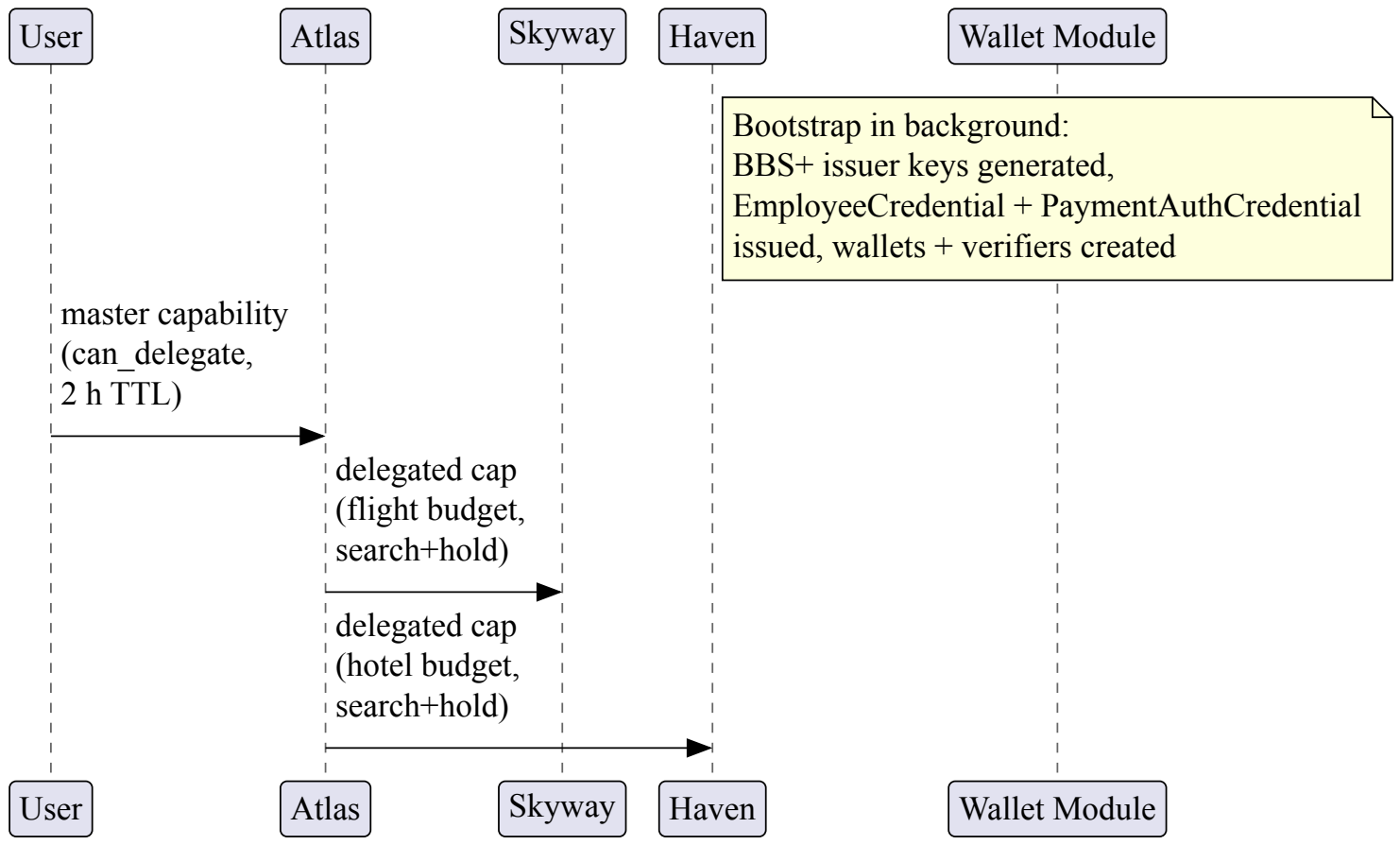

**Figure 5:** Phase 1 — Initialisation: wallet bootstrap and capability token issuance.

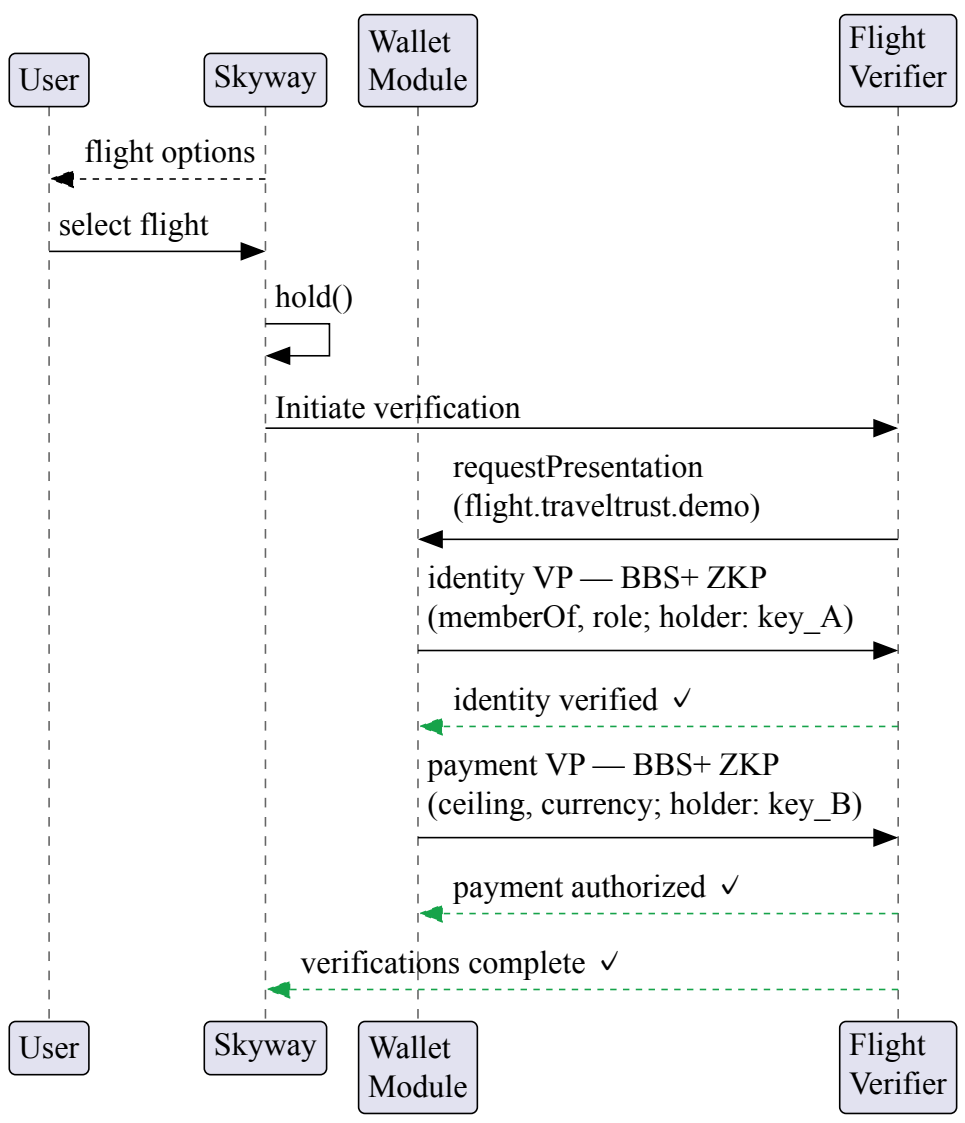

**Figure 6:** Phase 2 — Flight booking: Skyway holds a flight and the wallet presents BBS+ ZKPs to the flight service verifier.

During the flight booking process, shown in Figure 6 the wallet module is called on to present the user's credentials such as membership and budget availability via the BBS+ ZKP mechanism to the flight verifier. The verifier is expected to be under the control of the service responsible for committing the reservations whereas the agents merely facilitate the discovery of the possibilities and communicating the travel requirements.

The hotel booking phase shown in Figure 7 illustrates how the credential presentation to the hotel verifier is delinked from the flight verification process. A hotel might need to check that a user who is reserving their stay is to come under corporate booking or personal booking, for example, and therefore might call on the wallet module to present credentials establishing the user as an individual travelling for business reasons and the company that will be responsible for reimbursing the stay costs. This communication is purely between the wallet module and the hotel verifier and is merely initiated by the agents, as mentioned earlier.

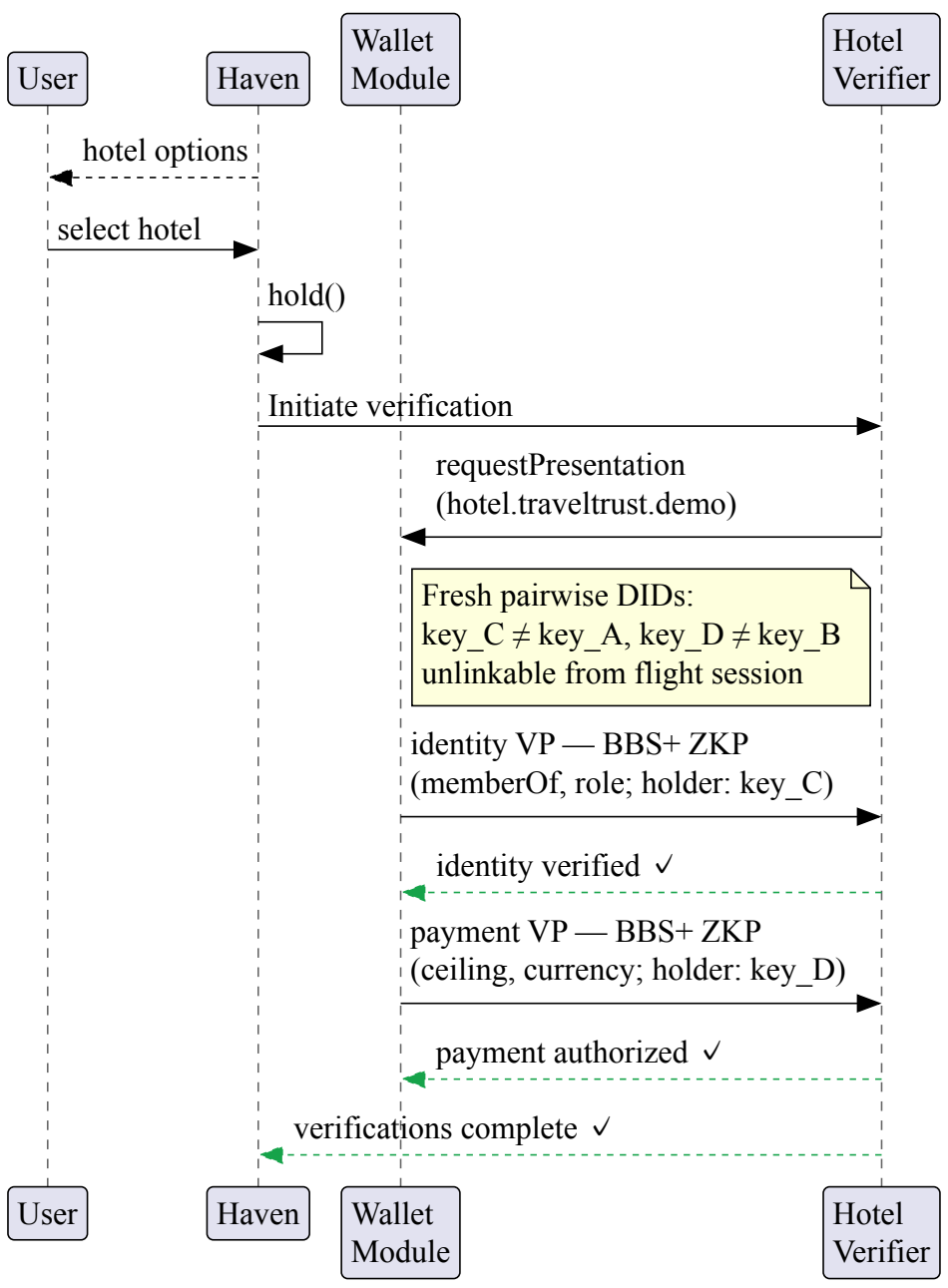

**Figure 7:** Phase 3 — Hotel booking: Haven holds a hotel and the wallet uses fresh pairwise DIDs (key_C, key_D) unlinkable from the flight session.

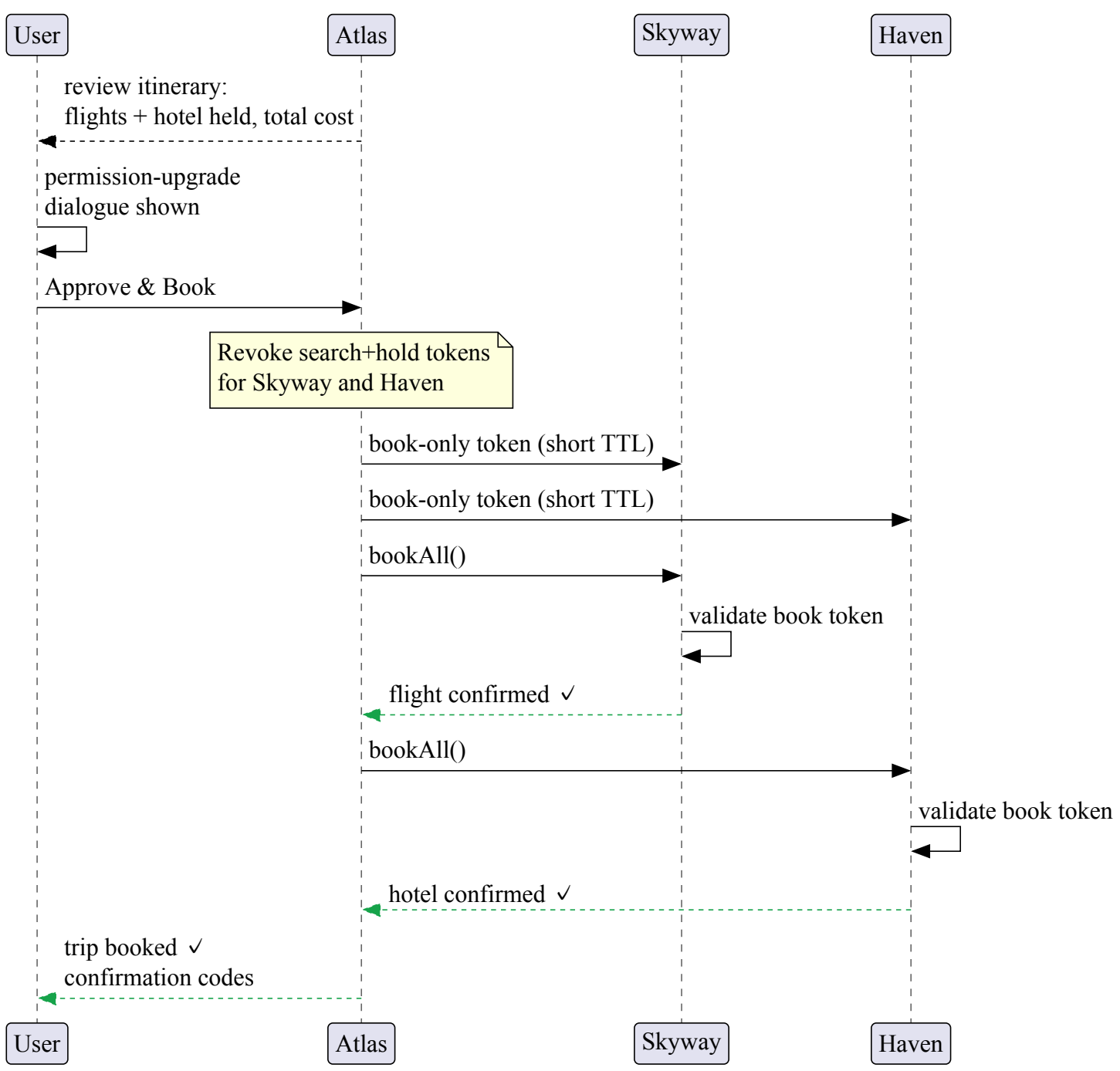

**Figure 8:** Phase 4 — Permission upgrade and final booking: search+hold tokens are revoked, book-only tokens are issued after user approval, and bookAll() executes.

Once the resources needed for the trip are confirmed by the user and a "go ahead" is issued, the "hold" tokens initially issued will need to be revoked and replaced with short lived "book" tokens. The respective services, "Skyway" and "Haven" in this case, then communicate the confirmation to the agents which eventually bubble up the confirmation to the user. The coordinator agent might also be expected to summarize the activity and present the spend breakdown, in an appropriate user interface.

A salient feature of this process is that the creation, exchange and revocation of tokens as part of hand overs between the agents and the services offers a cryptographically verifiable audit trail. Had the user been the only person to take each action, this audit trail would've been guaranteed to have a strict temporal ordering. However, since multiple agents are acting on behalf of the user towards the end result, the audit trail will have multiple linear-time "threads" guided by the sequencing of requests and verification presentations.

Figure 9 shows details of the two behind-the-scenes phases of credential presentation and payment presentation. While this process describes how the wallet module facilitates this, the final payment execution will have to be implemented using appropriate mechanisms by payment providers. The main purpose of these processes therefore is to offer a verification trail of all the decisions that went into the triggering of the final payment steps.

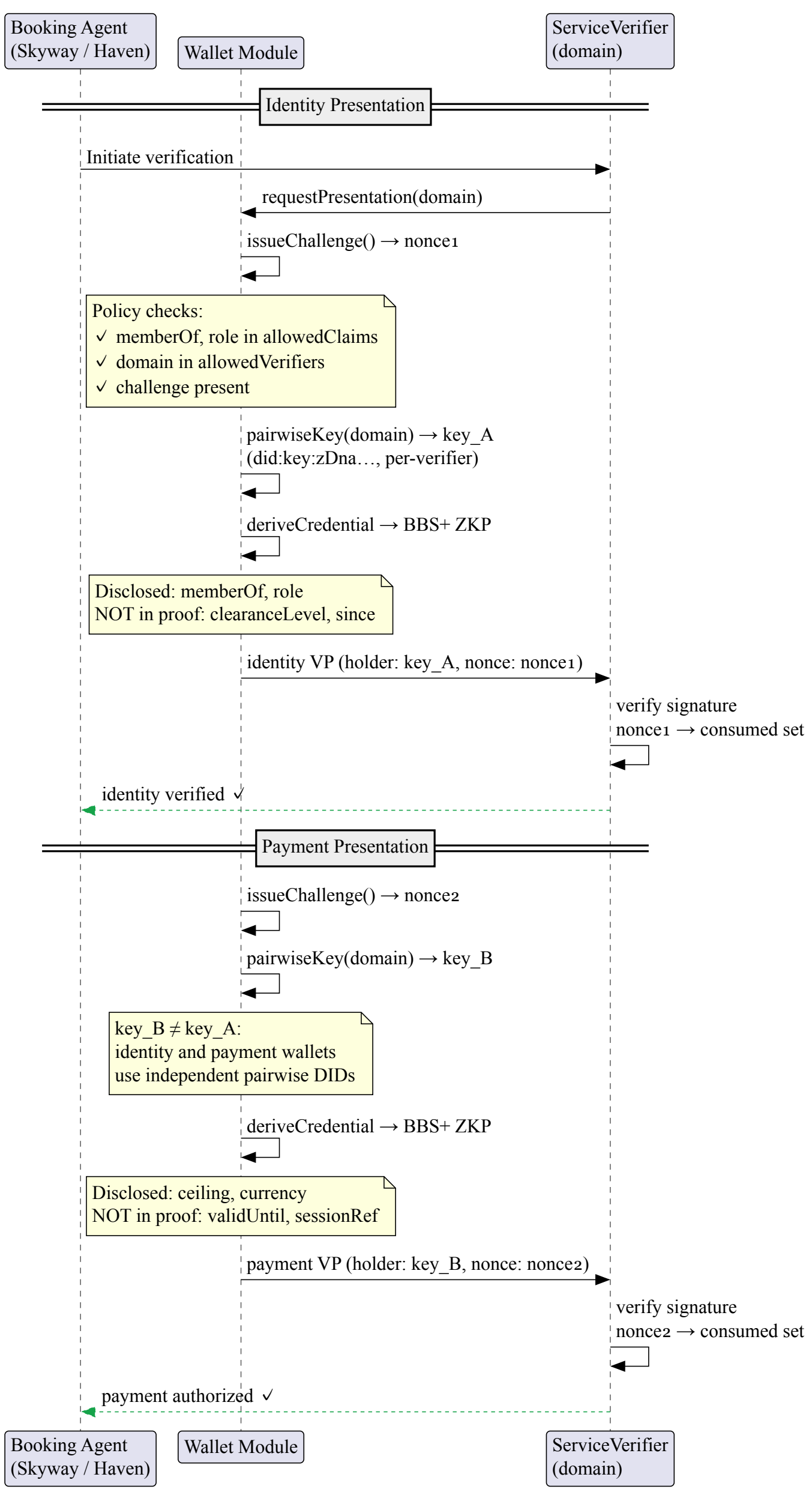

**Figure 9:** Credential presentation detail — wallet policy checks, BBS+ ZKP derivation, and nonce tracking for a single service.

In the entire flow, we expect agents to be accessed via described service end points hosted behind https access for link security. The agents therefore confirm their basic identity by showing control over the internet domain under which they're hosted. Subsequently, their claim over their DID and how it ties into the domain under which they're hosted further confirms that the right agent is being called to act. Agent discovery is not in scope for this proposal, as it can be implemented as a trusted discovery service such as a search engine, or even a known "discovery agent".

In an ecosystem where we expect multiple potentially ephemeral agents to be deployed in a decentralized or federated manner, such a discovery process also poses a potential attack surface where malicious agents can be injected into the discovery process. As with much of the internet, there is little a discovery service can do about this except via another source of trust comparable to domain certificate authorities.

Given that multiple agents may support similar services, there needs to be a way by which the coordinator (known as "Atlas" here) can choose the agent with appropriate abilities. We expect this will be done via a known protocol such as an OpenAPI spec located at a `/.well-known/` location[2] under the agent's domain. The role of the discovery agent therefore is to facilitate filtering down the set of available agents based on the required abilities without having to incur redundant LLM costs by processing the API specs repeatedly for each instance of usage. Such a discovery agent might build its repository by lazy-indexing agents over time and caching the index for use in the discovery phase.

[2] https://en.wikipedia.org/wiki/Well-known_URI

## CONCLUSION

Cryptographically bound, identity verified tokens with embedded constraint information that only attenuates with delegations are a scalable approach to managing the resource access of ad hoc triggered agents that act on behalf of users. The W3C DID and VC standards and the OAuth draft proposal for agents encapsulate these techniques in a standards path ripe for adoption when agents begin to take on. Additionally, the BBS-2023 mechanism can help improve privacy by permitting sharing of select credentials depending on the task at hand. In a world where such agents, enabled by LLM based functionality, can help users by collaborating on complex tasks, it is imperative to ensure that the architecture of these agents can provide guarantees about the actions they can take on behalf of users.